\documentclass[12pt]{article}

\usepackage{graphics}

\def\bhz{{\bf \hat z}}
\def\bht{{\bf \hat t}}

\begin{document}

\title{Effects of Kinks on DNA Elasticity}
\author{Yuri O.\ Popov\thanks{Corresponding author.  E-mail: {\tt yopopov@umich.edu}} \and Alexei V.\ Tkachenko}
\date{{\em Department of Physics, University of Michigan,} \\
{\em 500 E.\ University Ave., Ann Arbor, MI 48109}}
\maketitle

% \newpage

\begin{abstract}
We study the elastic response of a worm-like polymer chain with reversible kink-like structural defects.  This is a generic model for (a) the double-stranded DNA with sharp bends induced by binding of certain proteins, and (b) effects of trans-gauche rotations in the backbone of the single-stranded DNA.  The problem is solved both analytically and numerically by generalizing the well-known analogy to the Quantum Rotator.  In the small stretching force regime, we find that the persistence length is renormalized due to the presence of the kinks.  In the opposite regime, the response to the strong stretching is determined solely by the bare persistence length with exponential corrections due to the ``ideal gas of kinks''.  This high-force behavior changes significantly in the limit of high bending rigidity of the chain.  In that case, the leading corrections to the mechanical response are likely to be due to the formation of multi-kink structures, such as kink pairs.
\end{abstract}

\begin{center}
{\bf PACS}:  87.14.Gg, 87.15.La, 82.37.Rs, 87.15.Aa.
\end{center}

% \newpage

\section{Introduction}

Since the pioneering experiments of Smith {\em et al.}~\cite{smith1}-\cite{smith4}, the single-molecule micromechanical studies have become one of the central themes in molecular biophysics.  These techniques have contributed substantially into our understanding of the properties of various biomolecules and their interactions.  The original stretching experiments with the double-stranded DNA (ds-DNA) were followed by the theoretical work of Marko and Siggia~\cite{marko} who demonstrated that the observed elastic response ({\em i.e.}\ the dependence of the ds-DNA end-to-end extension on the applied stretching force) is in remarkable agreement with the Worm-Like Chain (WLC) model of polymer elasticity.  Within this model, the chain is described as a constant-length rod with finite bending rigidity subjected to the thermal fluctuations.  Various modifications and refinements of the WLC model have been proposed over the past decade to include additional effects, such as torsional constrains~\cite{moroz}, bond elasticity, and sequence disorder~\cite{nelson, bensimon}.  Recently, a discrete version of the WLC has been suggested as a plausible model for the single-stranded DNA (ss-DNA)~\cite{storm}.  In most cases, the modifications of the WLC model are within the domain of linear elasticity, and the overall non-linearity of the observed response is associated with the entropic nature of the problem.

In this paper, we discuss an intrinsically non-linear generalization of the WLC that takes into account effects of the localized structural defects on its elastic response.  In particular, we study a generic model of the WLC with reversible kink-like singularities.  This model is relevant to several systems of interest.  First, the kink-like defects can be used to model the sharp bending of the ds-DNA backbone associated with binding of certain proteins~\cite{ptashne}-\cite{thomas}.  In addition, the protein-mediated looping of the DNA would normally result in a non-trivial mutual alignment of the ends of the loop, which can also be interpreted as a kink defect.  Typically, such DNA-protein interactions are sequence-dependent, and there is a strong preference for the proteins to bind to the particular sites on the DNA.  However, in a realistic situation the non-specific binding takes place as well, and it may affect the overall properties of the DNA in vivo.  Since the stretching experiments can probe the elastic properties on the single-molecule scale, they may be used to extract the information about the parameters of such DNA-protein interactions.  The theoretical study of this paper is important for the interpretation of the results of this type of future experiments.  This problem was recently addressed in a general context of the DNA-deforming proteins~\cite{yan1}.  In that work, a discrete version of the WLC model was used to obtain the stretching curves with the numerical transfer matrix technique.  In our paper, use of the continuous model allows us to obtain a variational analytical result and compare it against the numerics.  In addition, our approach yields important insights into the underlying physics associated with the predicted elastic behavior.  The only significant limitation of the continuous model is that it does not allow one to approach the (hypothetical) saturation regime where the DNA gets completely covered by the proteins.

The other system where our model is relevant is the ss-DNA.  While several models have already been proposed to describe its elastic properties, neither of them is sufficiently justified by the microscopic structure of the molecule, and all of these models break down at high enough stretching forces.  Analysis of the backbone structure of the ss-DNA suggests that the traditional models such as the WLC or the Freely-Joint Chain (FJC) are unlikely to be adequate as the coarse-grained description of the chain.  Indeed, on the atomic level the conformations of the ss-DNA are mainly associated with rotations of certain bonds.  It is well known in the context of the general polymer physics that such rotations typically involve transitions between the discreet {\em trans\/} and {\em gauche\/} states, which are normally described with the Rotational Isomer model~\cite{flory}.  On the other hand, there are also elastic modes associated with small deviations of the torsional angles from their local equilibrium values ({\em i.e.}\ from these {\em trans\/} and {\em gauche\/} states).  The WLC would be a natural model for describing the long-wavelength elastic modes due to these small deformations.  In order to account for both types of conformations, one needs to construct a hybrid description that would unify the two classical models of polymer physics.  In fact, our problem of the WLC with reversible kinks has all the essential features of such a hybrid model.  Indeed, the local {\em trans\/} conformation can be viewed as a ``no-kink'' state, while the {\em gauche\/} state can be considered as a kink since it corresponds to a turn in the overall direction of the backbone.

Our model is different from a recent model by Wiggins, Phillips and P.~Nelson~\cite{wiggins}, where the kinks are taken as freely-bending hinges instead of the fixed-angle singularities considered here.  In most cases of the biological relevance the fixed-angle kinks appear as a more adequate description of the respective singularities than the free hinges.  In particular, the {\em gauche\/} states are characterized by the specific angles between the bonds fixed by the chemistry of the polymer (ss-DNA).  Similarly, protein binding and protein-mediated looping feature some characteristic angles determined by the nature of the proteins, although these angles may vary over some range of values~\cite{noort} (we consider modification of our fixed-angle results for the case of these deformable kinks with thermal fluctuations at the very end of this work).  Our results will be seen to depend on the kink angle quite substantially, and thus this extra parameter is important.  A model with soft annealed kinks similar to \cite{wiggins} was also considered recently by Yan and Marko~\cite{yan2} in the context of DNA cyclization.

In the following section we first describe our model for a worm-like chain with reversible kinks, and then outline a general approach to its solution by drawing an analogy to the Quantum Rotator.  Subsequently, we solve the resulting evolution equation both analytically (by the variational method) and numerically (by direct integration), and compare the results.  Finally, we discuss our results in both the weak-force and the strong-force limits as well as in the intermediate regime, and draw physical conclusions from our findings.

\section{Theory}

\begin{figure}
\begin{center}
\includegraphics{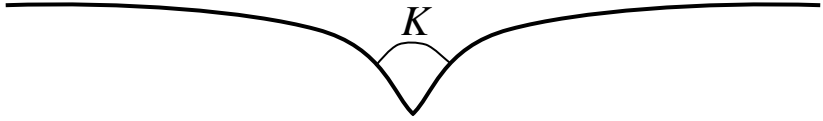}
\caption{Kink geometry.}
\label{geometry}
\end{center}
\end{figure}

\subparagraph{Model.}  We consider a worm-like polymer chain with rod-like bending elasticity.  The persistence length of the unperturbed chain is denoted by $l_p$, and the total length of the chain $L$ is much greater than the persistence length throughout this paper.  We presume that kink-like structural defects can exist anywhere along the chain.  These defects are reversible and can appear and disappear spontaneously, with certain free-energy penalty associated with each of them.  We should emphasize that we neglect the sequence dependence of kink probability, thus limiting ourselves to the case of non-specific DNA-protein binding only.  Each kink is characterized by the (fixed) opening angle $K$ (Fig.~\ref{geometry}); this value is the same for all the kinks and is an external parameter of the problem.  The presence of each kink costs finite amount of the free energy $\epsilon$ (at zero stretching force), which is also assumed to be the same for each kink.  The structural defects are local, {\em i.e.}\ they are characterized by a microscopic length scale $l_0$ much smaller than the persistence length $l_p$ and can be considered as point-like for most practical purposes.  We also neglect any direct interaction of the kinks separated by more than $l_0$.  Thus, the three length scales of the problem are related by $L \gg l_p \gg l_0$.

An important quantity is the average line density of the kinks $\kappa$ in the absence of the applied stretching force.  This quantity is defined as the Boltzmann probability of the existence of a kink per characteristic length scale of a kink site $l_0$:
\begin{equation}
\kappa = \frac{1}{l_0} \frac{\exp(-\epsilon/kT)}{1+\exp(-\epsilon/kT)}.
\end{equation}
At each site of length $l_0$, there might be either a kink [factor $\exp(-\epsilon/kT)$] or no kink [factor 1], with the probability of having a kink at an arbitrary site (per unit length) given by the expression above.  Note that in practice we assume that $\epsilon \gg kT$, and therefore
\begin{equation}
\kappa \approx \frac{1}{l_0} \exp(-\epsilon/kT) \ll \frac{1}{l_0},
\end{equation}
{\em i.e.}\ kinks are rare and far apart, and only few possible sites are occupied by the kinks.

The chain is stretched by applying force $F$ in the $\bhz$ direction of the Euclidian coordinates.  The effective energy of a chain segment between two adjacent kinks is given by the sum of the bending energy and the coupling to the stretching force:
\begin{equation}
\frac{E}{kT} = \int_{s_i}^{s_{i+1}} \left[ \frac{l_p}2 \left( \frac{\partial\bht}{\partial s} \right)^2 - \frac{F}{kT} \bhz\cdot\bht \right]\,ds.
\label{energy}
\end{equation}
Here $s$ is the coordinate along the chain, $\bht(s)$ is the unit tangent vector of the chain at $s$, and kink number $i$ is located at coordinate $s_i$ [so that $L = \sum_i (s_{i+1} - s_i)$].  This expression was used by Marko and Siggia~\cite{marko} and in some earlier works on worm-like chain elasticity.  This time, however, it does not apply to the entire chain; instead, it applies only to the chain segments between kinks.  At each kink ({\em i.e.}\ at $s_i$) the tangent vector experiences an abrupt change of its orientation and hops to a direction $\pi - K$ away from the preceding one.  Mathematically, this can be written as a constraint
\begin{equation}
\bht(s_i - 0)\cdot\bht(s_i + 0) = \cos(\pi - K) = - \cos K.
\label{constraint}
\end{equation}

\subparagraph{Analogy to the Quantum Rotator, and the free energy.}  Let $\psi(\bht,s)$ be the orientational distribution function for the chain ends.  Then the probability distribution for the tangent vector $\bht$ at coordinate $s$ for inner points of the chain is simply $|\psi(\bht,s)|^2$.  Function $\psi$ satisfies a Schrodinger-like equation for evolution along the chain.  This can be demonstrated by the path-integral technique~\cite{marko,fixman} where path integration is conducted over all possible chain conformations, thus accounting, among other things, for the entropy of the chain.  A more illustrative (though less rigorous) way to obtain the same equation also exists~\cite{strick}.  Here, we will only notice that each term in the effective energy~(\ref{energy}) has a counterpart in this Schrodinger-like equation, in the same way as each term of the mechanical energy in classical mechanics has a counterpart in the quantum-mechanical Hamiltonian.  Thus, an unperturbed chain (without any kinks) obeys the evolution equation $\partial_s \psi(\bht,s) = \hat H_0 \psi(\bht,s)$, where $\hat H_0$ is the effective Hamiltonian
\begin{equation}
\hat H_0 = \frac{1}{2l_p} \Delta_\bht + \frac{F}{kT} \bhz\cdot\bht.
\end{equation}
The term with the Laplacian in $\bht$ arises from the term with the $s$-derivative in the effective energy~(\ref{energy}), while coupling to the external force remains the same [up to the sign change] (see Refs.~\cite{marko,fixman} for details).  The Hamiltonian $\hat H_0$ is the Hamiltonian of the Quantum Rotator in quantum mechanics. 

Now we need to add a term to this Hamiltonian responsible for the non-local constraint~(\ref{constraint}).  This teleportation-like term changes the direction of $\bht$ abruptly, with the tangential vector $\bht(s)$ hopping instantly by a finite angle $\pi - K$ at the location of a kink.  Such behavior is described by a delta-functional kernel in the Hamiltonian:
\begin{equation}
\hat V \psi(\bht,s) \propto \int \frac{d^2\bht'}{2\pi} \delta(\bht\cdot\bht'+\cos K) \psi(\bht',s),
\label{v-term}
\end{equation}
where the full Hamiltonian is now $\hat H = \hat H_0 + \hat V$ and the $1/2\pi$ factor provides the proper normalization of the kernel.  The number of such non-local hops of the tangential vector ({\em i.e.}\ the number of kinks) yields the proportionality constant in the expression above.  Thus, the missing pre-factor is simply the average line density of the kinks $\kappa$, and the full Hamiltonian is 
\begin{equation}
\hat H \psi(\bht,s) = \frac{1}{2l_p} \Delta_\bht \psi(\bht,s) + \frac{F}{kT} \bhz\cdot\bht \psi(\bht,s) + \kappa \int \frac{d^2\bht'}{2\pi} \delta(\bht\cdot\bht'+\cos K) \psi(\bht',s).
\label{hamiltonian}
\end{equation}

The structure of the evolution equation
\begin{equation}
\frac{\partial\psi(\bht,s)}{\partial s} = \hat H \psi(\bht,s),
\label{evolution}
\end{equation}
suggests the standard procedure for solving this kind of quantum-mechanical-like problems:  expansion in eigenfunctions of the time-independent ({\em i.e.}\ $s$-independent) Schrodinger equation
\begin{equation}
- \mu \psi(\bht) = \hat H \psi(\bht),
\label{eigenproblem}
\end{equation}
where $\psi(\bht)$ and $\mu$ are the eigenfunctions and the eigenvalues, respectively.  Since each of the terms in such an expansion depends on the exponent of (the negative of) the corresponding eigenvalue, it is easy to conclude that for sufficiently long chains most terms in the expansion die off very fast, and only the smallest-eigenvalue term governs the long-chain behavior.  Thus, we are interested in the lowest eigenvalue $\mu$ of the equation~(\ref{eigenproblem}).  The free energy of the chain is then simply related to this lowest eigenvalue by
\begin{equation}
\frac{\cal F}{kT} = L \mu
\end{equation}
(see Refs.~\cite{marko,fixman} for details).

\begin{figure}
\begin{center}
\includegraphics{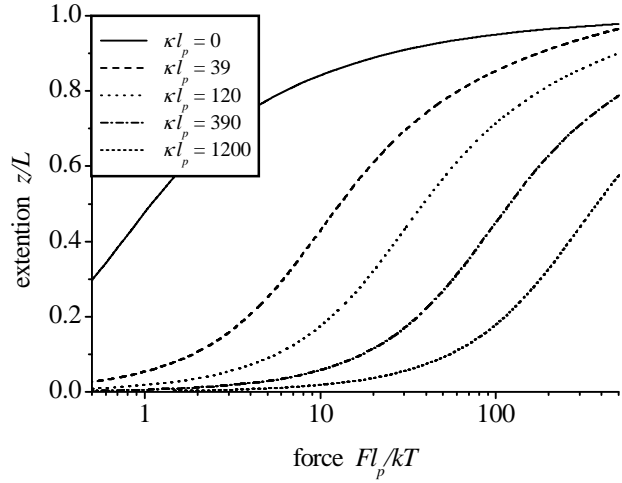}

(a)

\includegraphics{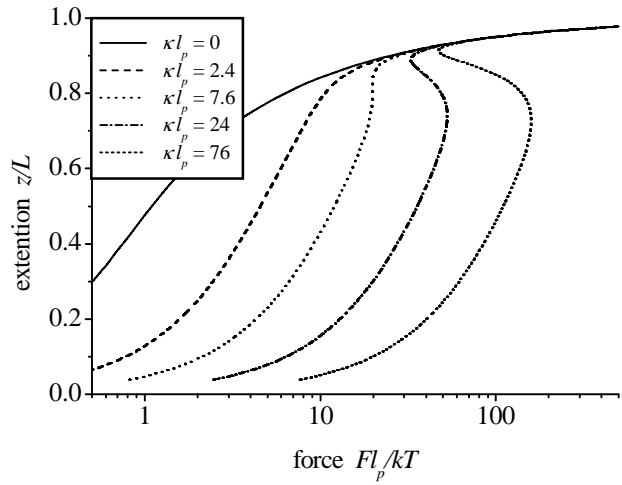}

(b)

\caption{Extension curves: force vs.\ extension.  Analytical results of the variational solution (a) for $K = 135^{\circ}$, and (b) for $K = 45^{\circ}$.  Curves for several values of the kink density $\kappa l_p$ are shown on each plot.}
\label{extensioncurve}
\end{center}
\end{figure}

\subparagraph{Variational solution.}  Since the lowest eigenvalue is being sought, we use the variational method for finding the analytical solution to eigenproblem~(\ref{eigenproblem}).  Application of the force makes the $\bhz$ direction highly preferable for the tangential vector, and therefore the distribution function $\psi(\bht)$ must be strongly-peaked at $\bht \| \bhz$.  Conventionally, we choose $\psi(\bht) \propto \exp (\omega \bhz\cdot\bht /2)$ as a trial function, or
\begin{equation}
\psi(\bht) = \sqrt{\frac\omega{4\pi\sinh\omega}} \exp\left(\frac{\omega \bhz\cdot\bht}2\right)
\label{trialfunction}
\end{equation}
upon proper normalization ($\int \psi^2(\bht)\,d^2\bht = 1$).  Here $\omega$ is the variational parameter.  The lowest eigenvalue can then be estimated as
\begin{equation}
\mu = - \max_\omega \bar H,
\end{equation}
where
\begin{equation}
\bar H = \int \psi(\bht) \hat H \psi(\bht) \, d^2\bht.
\end{equation}
Evaluating the last expression with trial function~(\ref{trialfunction}) yields
\begin{equation}
\mu = \min_\omega \left[\left(\frac\omega{4l_p} - \frac{F}{kT}\right) \left(\coth\omega - \frac{1}\omega\right) - \kappa \frac{\sinh(\alpha\omega)}{\alpha\sinh\omega}\right],
\label{freeenergy}
\end{equation}
where we introduced $\alpha = \sin(K/2)$.  Minimization with respect to $\omega$ gives the implicit dependence of $\omega$ on the applied force or, being inverted, the explicit dependence of the force on $\omega$:
\begin{equation}
\frac{F}{kT} = \frac{\omega^2}{4l_p} \frac{\cosh\omega \sinh\omega - \omega}{\sinh^2\omega - \omega^2} + \kappa \omega^2 \frac{\alpha^{-1} \sinh(\alpha\omega) \cosh\omega - \cosh(\alpha\omega) \sinh\omega}{\sinh^2\omega - \omega^2}.
\label{force}
\end{equation}
Note that here and everywhere below $\omega$ is the particular value of the variational parameter that minimizes the free energy (instead of the generic variable implied until this point).  Knowledge of the free energy as a function of force~(\ref{freeenergy}) also allows one to determine the extension of the chain:
\begin{equation}
\frac{z}L = - \frac{1}L \frac{\partial\cal F}{\partial F} = - \frac{\partial\mu}{\partial(F/kT)} = \coth\omega - \frac{1}\omega.
\label{extension}
\end{equation}
Thus, equations~(\ref{force}) and (\ref{extension}) provide the parametric dependence of the force on the extension, with $\omega$ playing the role of the parameter.  This extension curve is shown in Fig.~\ref{extensioncurve} for kink angles of $135^{\circ}$ and $45^{\circ}$ and several values of $\kappa$.  Our results reduce to those of Marko and Siggia~\cite{marko} when no kinks are present, {\em i.e.}\ when $\kappa = 0$.  The new physics introduced by the kinks is represented by the term proportional to $\kappa$ in equation~(\ref{force}).  The ratio of the first and the second terms in this equation is determined by the dimensionless parameter $\kappa l_p$, which has the physical meaning of the average number of kinks per the persistence length in the absence of force.  The larger this parameter is, the more significant the kink contribution is.

\subparagraph{Weak forces and small extensions.}  The asymptotics of the above results in the limit of small $\omega$:
\begin{equation}
\frac{z}L = \frac{\omega}3 + O\left(\omega^3\right)
\end{equation}
and
\begin{equation}
\frac{F}{kT} = \left[\frac{1}{2l_p} + \kappa (1 + \alpha^2) \right] \omega + O\left(\omega^3\right)
\end{equation}
allow one to obtain the explicit dependence of the extension on the force in the limit of weak forces and small extensions:
\begin{equation}
\frac{z}L = \frac{2}3 \frac{l_p}{1 + 2 \kappa l_p (1 - \alpha^2)} \frac{F}{kT}.
\end{equation}
This expression can be compared to the weak-force result for a bare (kink-free) persistence chain $z/L = (2/3) l_p (F/kT)$.  Clearly, the elastic response of the chain with kinks is characterized by the renormalized persistent length
\begin{equation}
l_{p\,{\rm eff}} = \frac{l_p}{1 + 2 \kappa l_p (1 - \alpha^2)} = \frac{l_p}{1 + \kappa l_p (1 + \cos K)}.
\end{equation}
This quantity reduces to the bare persistence length $l_p$ in the absence of kinks ($\kappa = 0$).  It is also apparent that $l_{p\,{\rm eff}} < l_p$, and therefore a worm-like chain with kinks is {\em harder\/} than a worm-like chain without them.  Similar renormalization of the persistence length was also observed in recent numerical results by Yan and Marco~\cite{yan1}.  It may appear that analogous renormalization was obtained in Refs.~\cite{nelson, bensimon} as well, however, these works consider quenched sequence disorder, and thus the physical origin of their renormalization is very different.

\begin{figure}
\begin{center}
\includegraphics{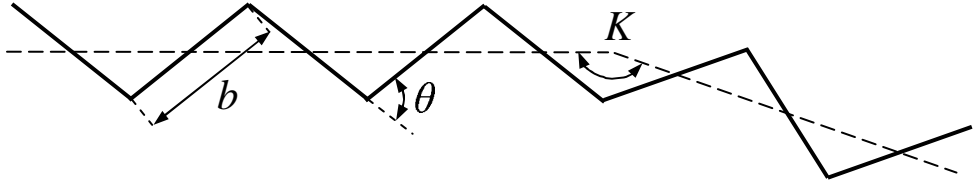}
\caption{Geometrical parameters of the Flory model for {\em trans-gauche\/} rotational isomers~\cite{flory}.}
\label{zigzag}
\end{center}
\end{figure}

For high bending rigidity ($\kappa l_p \gg 1$), the renormalized persistence length becomes $ l_{p\,{\rm eff}} = 1/[\kappa (1 + \cos K)]$.  This result deserves some discussion as it turns out to be closely related to the Flory result for {\em trans-gauche\/} rotational isomers~\cite{flory}.  If a {\em gauche\/} conformation is considered as a kink, then it is straightforward to identify the geometrical relations
\begin{equation}
\cos\frac{K}2 = \frac{\sqrt{3}}2 \sin\frac{\theta}2
\label{correspondence}
\end{equation}
and $l_0 = b \cos(\theta/2)$, where $\theta$ is the angle characterizing the {\em trans\/} zigzag and $b$ is the length of the zigzag segment (Fig.~\ref{zigzag}).  From these results, one can obtain the mean square of the distance between the chain ends:
\begin{equation}
\langle R^2 \rangle = 2 l_{p\,{\rm eff}} L = 2 \frac{l_{p\,{\rm eff}}}{l_0} N l_0^2 = 2 \frac{2 + \exp\left(\frac\epsilon{kT}\right)}{3 \sin^2\left(\frac{\theta}2\right)} N b^2 \cos^2\frac{\theta}2,
\end{equation}
which is identical to the result for the Flory model~\cite{flory}
\begin{equation}
\langle R^2 \rangle = N b^2 \frac{1+\cos\theta}{1-\cos\theta} \frac{1 + 2 \exp\left(\frac{\epsilon}{kT}\right)}3
\end{equation}
in the limit of rare kinks ($\epsilon \gg kT$).  Note that comparison to the Flory result for the free rotational model (instead of the {\em trans-gauche\/} rotational one) would be impossible due to the lack of the equivalent angle and absence of a relation similar to Eq.~(\ref{correspondence}).  In the case of the free rotations, the effective kink angle can adopt a continuous spectrum of values between zero and some positive value corresponding to the rotation by $180^{\circ}$.

\subparagraph{Strong forces and large extensions.}  In the opposite limit of large $\omega$ our results~(\ref{force}) and (\ref{extension}) also allow for simple asymptotic expansions:
\begin{equation}
\frac{z}L = 1 - \frac{1}\omega
\end{equation}
and
\begin{equation}
\frac{F}{kT} = \frac{\omega^2}{4 l_p} + \kappa \omega^2 \frac{1-\alpha}\alpha \exp\left[-(1-\alpha)\omega\right].
\label{strongforces}\end{equation} 
Thus, an explicit dependence for the elastic response is again available:
\begin{equation}
\frac{z}L = 1 - \sqrt{\frac{kT}{F l_p}} \left[ \frac{1}2 + \kappa l_p \frac{1-\alpha}\alpha \exp\left(- 2(1-\alpha) \sqrt{\frac{F l_p}{kT}} \right) \right].
\end{equation}
To the main order, this result reproduces the earlier result of Marko and Siggia~\cite{marko} for the one-over-the-square-root dependence on the stretching force.  Hence, the response of a worm-like chain with kinks to the strong stretching is dominated mostly by the bare worm-like-chain elasticity.  In addition, kinks contribute an exponential correction to that main-order result, which can be interpreted as due to the ``ideal gas of kinks''.  Indeed, the second term in equation~(\ref{freeenergy}) is the average line density of the kinks in the presence of the stretching force, and it is equivalent to the concentration of the molecules in the ideal gas.  Thus, the second term in equation~(\ref{strongforces}) derived from the free energy~(\ref{freeenergy}) is analogous to the relation $p = n k T$ for the pressure of the ideal gas.

\begin{figure}
\begin{center}
\includegraphics{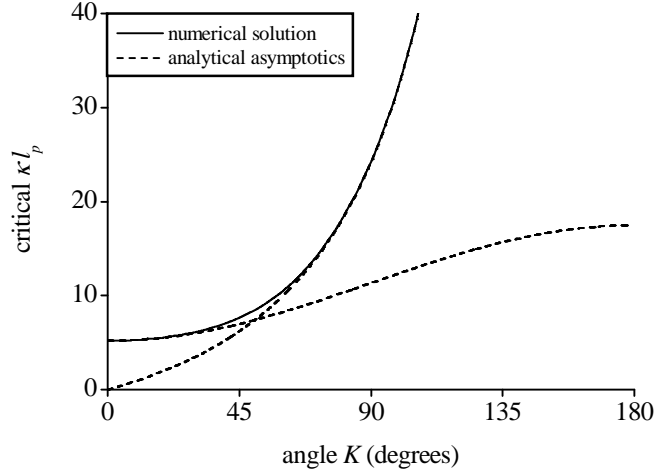}
\caption{``Critical'' value of parameter $\kappa l_p$ as a function of the kink angle $K$.  Both the numerical solution and the analytical asymptotics are shown.  The instability region is above the solid curve.}
\label{critical}
\end{center}
\end{figure}

\subparagraph{High rigidity, small kink angle, numerical results.}  Probably the most interesting feature of Fig.~\ref{extensioncurve} is related to the middle portion of the high-rigidity extension curves, where an instability region with $dz/dF < 0$ is present.  This instability region exists when parameter $\kappa l_p$ is greater than the ``critical'' value $(\kappa l_p)_c$, {\em i.e.}\ when the chains are of high rigidity.  The ``critical'' value is determined by the conditions $F'(z) = F''(z) = 0$ and is given by
\begin{equation}
(\kappa l_p)_c = \frac{\exp\omega_c}{\frac{\sinh\alpha\omega_c}{\alpha}\left(4\omega_c\left(1 + \alpha^2\right) - 8\right) + \cosh\alpha\omega_c\left(8 - 8\omega_c\right)},
\end{equation}
where $\omega_c$ is the solution to
\begin{equation}
\frac{\tanh\alpha\omega}{\alpha} = \frac{\omega\left(3 + \alpha^2\right) - 6}{\omega\left(1 + 3\alpha^2\right) - \left(3 + 3\alpha^2\right)}
\end{equation}
and is large for all kink angles (the lowest $\omega_c$ is around 5 for small kink angles).  The last equation allows for simple analytical solutions in the limits of small kink angles ($\alpha\omega \ll 1$) and large kink angles ($\alpha\omega \gg 1$).  Thus, the analytical asymptotics of $(\kappa l_p)_c$ in these two limits are readily available:
\begin{equation}
(\kappa l_p)_c = \frac{\exp\left(3 + \sqrt{3}\right)}{8\left(1 + \sqrt{3}\right)} \left(1 + 3 \alpha^2 \frac{2 + \sqrt{3}}{3 + \sqrt{3}}\right)\qquad\qquad\left(\alpha\omega \ll 1\right)
\end{equation}
and
\begin{equation}
(\kappa l_p)_c = \frac{\alpha \exp 3}{2(1 - \alpha)}\qquad\qquad\left(\alpha\omega \gg 1\right).
\end{equation}
These asymptotics as well as the exact numerical solution for $(\kappa l_p)_c$ are shown in Fig.~\ref{critical}.  As can be observed in this figure, the ``critical'' value is much lower for sharp kinks, and hence for smaller kink angles the instability region is present for much lower values of $\kappa l_p$ and lies in the range of the measurable stretching forces in Fig.~\ref{extensioncurve}.

This instability behavior is similar to the behavior of the Van-der-Vaals curves for the ordinary liquid-gas system and hence suggests the co-existence of two ``phases''.  Of course, no true phases can exist in a one-dimensional system like the DNA molecule under consideration, however, for the worm-like chain with kinks it may be appropriate to think of ``kink-free'' and ``kink-rich'' portions of the chain instead of the true phases.

In order to check this ``two-phase'' hypothesis, we conduct a direct numerical solution of the evolution equation~(\ref{evolution}) with the Hamiltonian~(\ref{hamiltonian}).  In order to avoid the delta-functional singularity inside the integration in the last term, the analytical integration of the delta-function is carried out, and we operate with a smooth-kernel Hamiltonian
$$\hat H \psi(p,s) = \frac{1}{2l_p} \left[ \left( 1 - p^2 \right) \frac{\partial^2 \psi(p,s)}{\partial p^2} - 2 p \frac{\partial \psi(p,s)}{\partial p} \right] + \frac{F}{kT} p \psi(p,s) \, +$$

\begin{equation}
+ \, \frac{\kappa}{\pi} \int^{-p \cos K + \sqrt{1-p^2} \sin K}_{-p \cos K - \sqrt{1-p^2} \sin K} \frac{\psi(p',s)\,dp'}{\sqrt{\left( 1 - p^2 \right) \left( 1 - p'{}^2 \right) - \left( p p' + \cos K \right)^2}}
\end{equation}
(where $p = \bhz\cdot\bht$ and $p' = \bhz\cdot\bht'$) instead of the original Hamiltonian~(\ref{hamiltonian}).  Then we choose an arbitrary initial distribution function $\psi(\bht,0)$ and let it evolve by evaluating the right-hand side of Eq.~(\ref{evolution}) on each step and setting the $s$-derivative (the left-hand side) equal to the result.  This procedure is stopped when the relative change of the relative extension $z/L = \int \bhz\cdot\bht \,\psi^2(\bht,s)\,d^2\bht$ evaluated at each step becomes very small and hence $\psi(\bht,s)$ itself becomes the stationary distribution function.  Such an approach yields more precise results for all the quantities since it does not make any {\em a priory\/} assumptions about the shape of the distribution function.

\begin{figure}
\begin{center}
\includegraphics{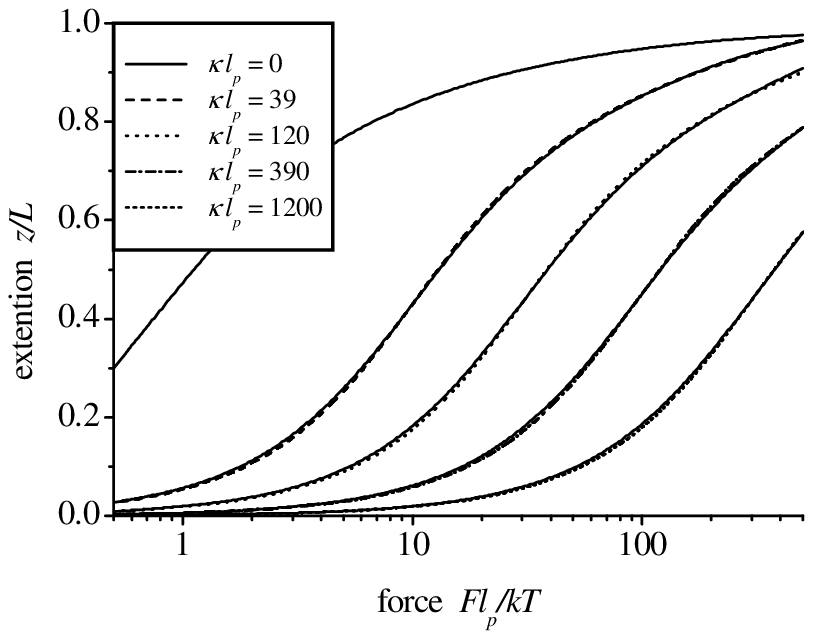}
(a)
\includegraphics{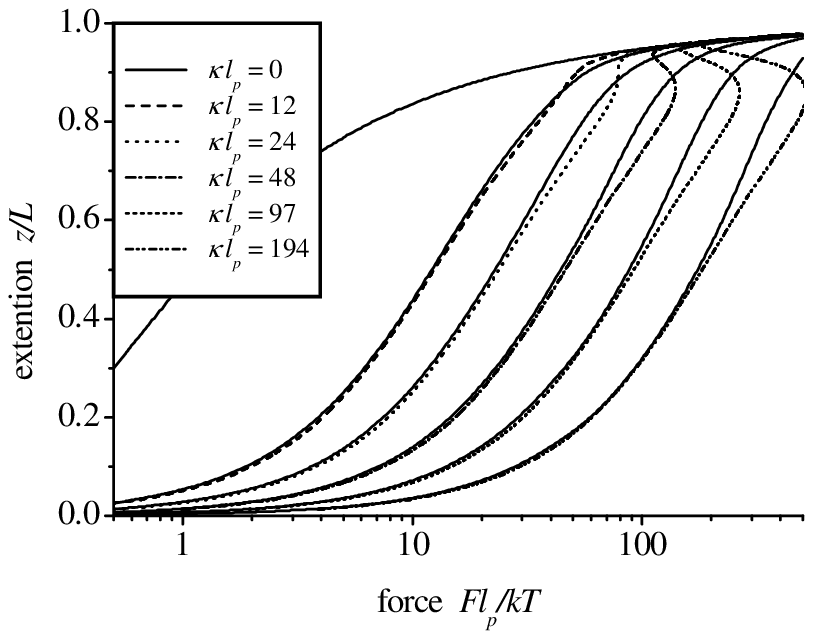}
(b)
\includegraphics{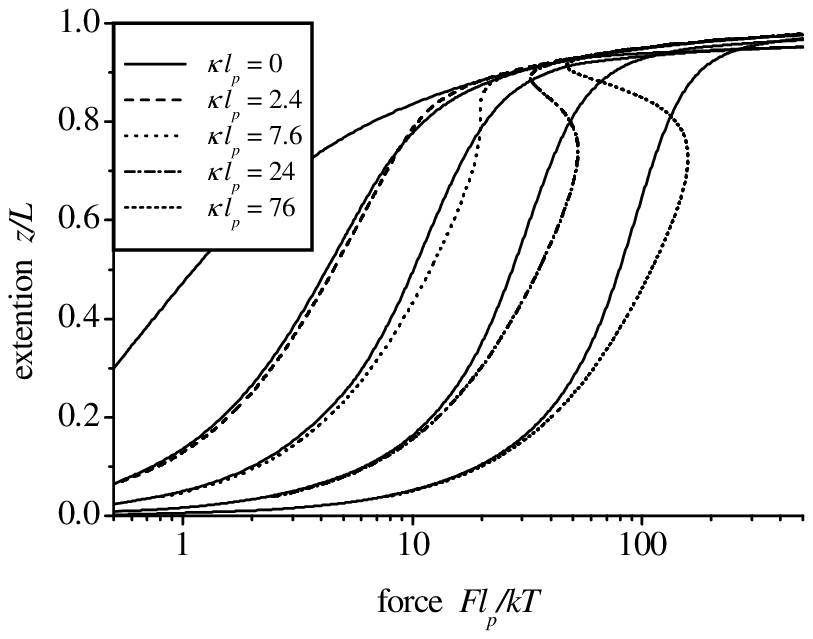}
(c)
\caption{Extension curves: force vs.\ extension.  Numerical results of the direct solution of the evolution equation (a) for $K = 135^{\circ}$, (b) for $K = 90^{\circ}$, and (c) for $K = 45^{\circ}$.  Curves for several values of the kink density $\kappa l_p$ are shown on each plot (same values as in Fig.~\ref{extensioncurve}).  Solid lines represent the numerical results, while the dashed lines reproduce the corresponding analytical results.}
\label{numericalresults}
\end{center}
\end{figure}

The results of this numerical solution are shown in Fig.~\ref{numericalresults} for kink angles of $135^{\circ}$, $90^{\circ}$, and $45^{\circ}$ and several values of $\kappa$.  It is immediately apparent that the agreement between the variational analytical and the exact numerical curves is good for both small and large forces (and extensions) for all values of parameters, and hence the analytical results of the preceding two sections are accurate in the respective limits.  However, the intermediate regime for high-rigidity chains and small kink angles differs substantially from the analytical parametric dependence of equations~(\ref{force}) and (\ref{extension}).  What causes this difference, and how can the numerical results be understood?

\begin{figure}
\begin{center}
\includegraphics{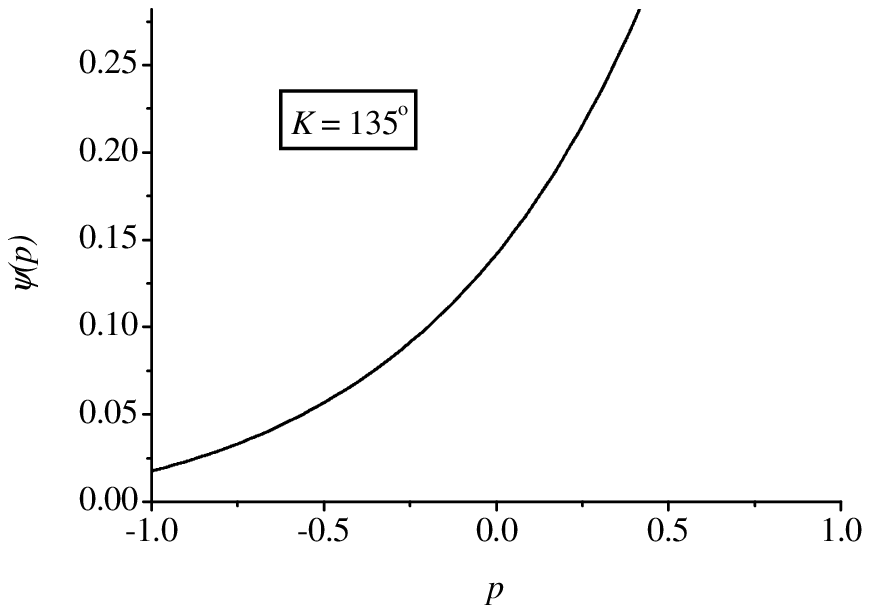}

(a)

\includegraphics{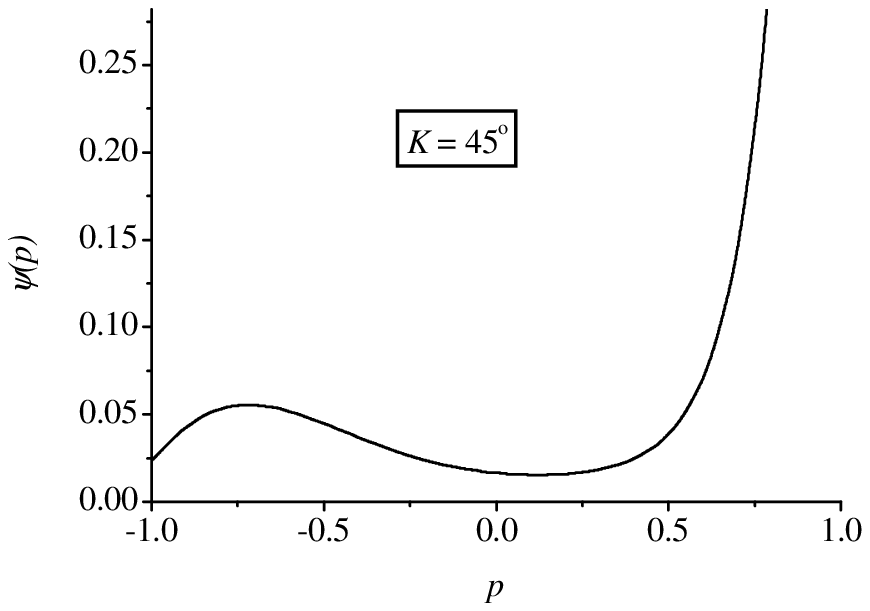}

(b)

\caption{Numerical results for the distribution function $\psi(p,s)$ (where $p = \bhz\cdot\bht$) in the limit of large $s$ for the kink angles of (a) $135^{\circ}$ and (b) $45^{\circ}$.  The two graphs are for the comparable stretching force and for the respective values of the kink density [$\sqrt{10}$ times the ``critical'' density for each kink angle, {\em i.e.}\ for the second from the right curve in Figs.~\ref{numericalresults}(a) and \ref{numericalresults}(c)].}
\label{distribution}
\end{center}
\end{figure}

\begin{figure}
\begin{center}
\includegraphics{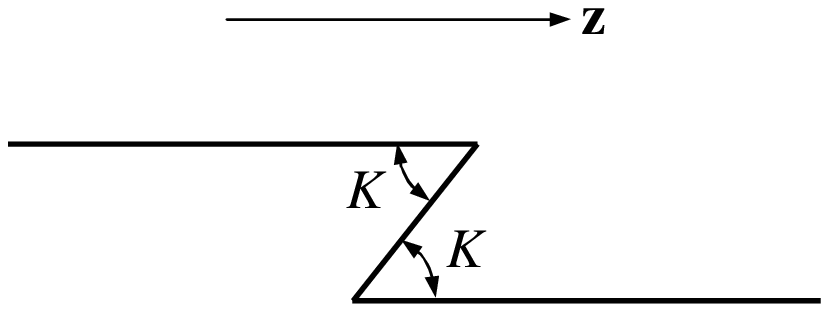}
\caption{A pair of kinks.}
\label{kinkpair}
\end{center}
\end{figure}

The answer to both these questions comes from the numerical results for the distribution function $\psi(\bht,s)$ (Fig.~\ref{distribution}).  The distribution profile for large kink angles possesses the shape assumed in the variational method: a peak at $\bht$ along $\bhz$ with the exponential fall-off away from it.  However, for rigid chains and small kink angles, we observe existence of a secondary peak in addition to the main peak.  Thus, the discrepancy between the variational and the numerical results can be explained by inadequacy of the trial function in the variational method, which ignores the possibility of the secondary maximum.  The location of the secondary peak corresponds approximately to the angle $\pi - K$ away from the direction of the applied force ($\bhz$), {\em i.e.}\ there is an unusually high portion of chain segments at angle $\pi - K$ to the force.  This can be interpreted as presence of a small, but substantial fraction of kink {\em pairs\/} (Fig.~\ref{kinkpair}).  Indeed, if the two outer chain sections of a kink pair are aligned precisely along $\bhz$, then the inner section is at angle $\pi - K$ to that direction, and if the kinks in each pair are close, then the fraction of the chain segments at angle $\pi - K$ to the force is small compared to the fraction of the chain segments aligned with the force, but high compared to the fraction of the other non-aligned segments [in full agreement with Fig.~\ref{distribution}(b)].  When the two kinks are in a pair with the outer sections aligned with the force and the inner section at angle $\pi - K$ to the force (Fig.~\ref{kinkpair}), much less bending is necessary compared to the case of a single kink with both outer sections aligned with the force (Fig.~\ref{geometry}).  Thus, condensation of kinks into pairs is favorable both because the fraction of the ``non-aligned with the force'' chain segments is small and because the energetically costly bending is not required.  We should emphasize that although kink-pairing may be a good way of thinking of this system, we do not have a simple yet rigorous theoretical model accounting for all the properties of such a ``gas of kink pairs''.

\subparagraph{Soft kinks.}  So far, we considered only hard kinks, {\em i.e.}\ kinks characterized by the fixed opening angle $K$.  Each kink could appear and disappear; however, the possibility of its elastic deformation due to the thermal fluctuations was ignored.  In this section, we demonstrate how our results are modified when kinks are of finite rigidity and their opening angle is allowed to deviate from the equilibrium value $K$.  We limit our discussion to small fluctuations around $K$, which corresponds to high rigidity of each kink.  In the opposite limit of large deviations, the kinks are almost free and the model of Wiggins {\em et al.}~\cite{wiggins} should be applied.

The effective energy of the chain can be written as a sum of the contributions~(\ref{energy}) of all segments between kinks and the quadratic contributions of all kinks:
\begin{equation}
\frac{E_{\rm soft}}{kT} = \sum_i \left\{ \int_{s_i}^{s_{i+1}} \left[ \frac{l_p}2 \left( \frac{\partial\bht}{\partial s} \right)^2 - \frac{F}{kT} \bhz\cdot\bht \right]\,ds + \frac{\gamma}2 \left[ \arccos\left(-\bht(s_i - 0)\cdot\bht(s_i + 0)\right) - K \right]^2 \right\},
\end{equation}
where the summation is over kinks $i$.  The last spring-like term allows for the thermal fluctuations of the opening angle and substitutes for the delta-functional constraint~(\ref{constraint}); a similar term was used in the numerical study by Yan and Marko~\cite{yan1}.  Kink stiffness parameter $\gamma$ is assumed to be large, so that the fluctuations are small.  The $\hat V$-term in the effective Hamiltonian is then
\begin{equation}
\hat V_{\rm soft} \psi(\bht,s) = \kappa \frac{\int e^{ - \frac{\gamma}2 \left[ \arccos\left(-\bht\cdot\bht'\right) - K \right]^2} \psi(\bht',s) \, d^2\bht'}{\int e^{ - \frac{\gamma}2 \left[ \arccos\left(-\bht\cdot\bht'\right) - K \right]^2} \, d^2\bht'} = \frac{\int e^{- \frac{\gamma}2 (K'-K)^2} \, \hat V \psi(\bht,s) \, d\cos K'}{\int e^{- \frac{\gamma}2 (K'-K)^2} \, d\cos K'},
\end{equation}
where in the last equality $\hat V$ is the old delta-functional term~(\ref{v-term}) (multiplied by $\kappa$) considered as a function of the integration variable $K'$ instead of the equilibrium value $K$.

Thus, the variational solution for the free energy of the chain with soft kinks reduces to a single integration of the last term of the variational result~(\ref{freeenergy}) for the free energy of the chain with hard kinks.  This integration can be performed analytically in the limit of small thermal fluctuations ($\gamma \gg \omega / \alpha$) and leads to the following main order result:
\begin{equation}
\mu_{\rm soft} = \min_\omega \left[\left(\frac\omega{4l_p} - \frac{F}{kT}\right) \left(\coth\omega - \frac{1}\omega\right) - \kappa \frac{\sinh(\alpha\omega)}{\alpha\sinh\omega} \exp\left(\frac{\omega^2(1-\alpha^2)}{8\gamma}\right)\right],
\end{equation}
or, in terms of the force,
\begin{equation}
\frac{F_{\rm soft}}{kT} = \frac{\omega^2}{4l_p} \frac{\cosh\omega \sinh\omega - \omega}{\sinh^2\omega - \omega^2} + \kappa \omega^2 \frac{\alpha^{-1} \sinh(\alpha\omega) \cosh\omega - \cosh(\alpha\omega) \sinh\omega}{\sinh^2\omega - \omega^2} \exp\left(\frac{\omega^2(1-\alpha^2)}{8\gamma}\right).
\end{equation}
In the expressions above we omitted all the terms of the order of $\omega/\gamma$ or less and retained only the term of the order of $\omega^2/\gamma$, which is not necessarily small when $\omega$ is large.  Thus, the main effect is simply renormalization of the kink density $\kappa$ by a factor of $\exp\left[\omega^2(1-\alpha^2)/(8\gamma)\right]$.  Physically, when kinks are allowed to relax by deforming the opening angle, the number of kinks increases, making the chain harder to extend.  However, the absolute value of the effect is not substantial, and for all reasonably small thermal fluctuations of the kink angle (large $\gamma$) the analytical extension curves get only slightly distorted for large extensions, without major qualitative changes.  For very soft kinks (small $\gamma$) the results of Wiggins {\em et al.}~\cite{wiggins} should be employed.

\section{Conclusions}

In conclusion, we studied the generic model of the semi-flexible polymer chain with reversible kinks.  It can be viewed as a hybrid of the two classical descriptions of the polymer elasticity:  the WLC and the rotational-isomer-states models.  Therefore, the proposed theory should be applicable to the ss-DNA whose conformations may involve both the discrete {\em trans-gauche\/} rotations of the chemical bonds and small deformations giving rise to the continuous elasticity.  Another important class of systems where our model is applicable is the ds-DNA with sharp protein-induced bends.

In the limit of weak stretching forces, the elastic response of the kinked DNA chain is characterized by a renormalized persistence length, which is smaller than the bare persistence length.  This conclusion is consistent with the observation made in the earlier numerical work on the problem~\cite{yan1}.  We obtained the analytical expression for the renormalized persistence length and showed that the classical results for both the pure WLC and the rotational isomer models can be recovered exactly as limits of our expression.  Furthermore, by using the variational approach, we calculated the complete non-linear response of the chain to the stretching.  This result is in excellent agreement with the direct numerical solution over a substantial range of the model parameters.

In the limit of strong stretching forces, we recover the pure worm-like-chain behavior with exponential corrections due to the ``ideal gas'' of kinks.  The variational theory breaks down in the regimes of high chain rigidity and small kink angles.  In this case, the analytical curves have signatures of instability similar to those of the Van~der~Waals gas.  By analyzing our numerical results, we conclude that this behavior corresponds to the creation of multi-kink objects, {\em e.g.}\ kink pairs.

For soft kinks, where the opening angle can fluctuate and thus relax the overall energy of the chain, we found that the number of kinks increases compared to the hard-kink case, and the DNA gets harder to extend.

The major limitation of our model is that we have neglected the sequence-specific effects by assuming that the kink energy is constant along the chain.  While this may be a reasonable first approximation to both problems of the ss-DNA elasticity and the non-specific protein-DNA binding, a significant future work is needed in order to include the effects of the sequence disorder.

Several experimental tests of our results can be suggested.  In the case of the ss-DNA, in order to probe the ``pure'' elastic response of the chain one needs to exclude the effects of base-pairing and the electrostatic interactions.  In the existing experiments, these effects are not suppressed, and thus direct comparison is not possible at the moment.  (Note that the electrostatic and the base-pairing contributions were {\em simulated\/} in Ref.~\cite{dessinges} instead of being excluded experimentally.)  However, in the future an experiment can be done at the conditions of very strong screening ({\em i.e.}\ at high salt concentration) with all-purine or all-pyrimidine ss-DNA sequences to avoid these contributions.

The effects of the protein-induced kinks on the ds-DNA elasticity can be studied by performing the DNA-stretching experiment at various concentrations of the DNA-binding proteins.

\vspace{3ex}

{\small  The authors acknowledge valuable discussions with J.-C.~Meiners, S.~Blumberg, Y.~Rabin, J.~Kierfeld, and T.A.~Witten.}

% \newpage

\end{document}